# Molecular Gas in the Halo Fuels the Growth of a Massive Cluster Galaxy at High Redshift

B. H. C. Emonts[1,*], M. D. Lehnert[2], M. Villar-Martín[1,3], R. P. Norris[4,5], R. D. Ekers[4], G. A. van Moorsel[6], H. Dannerbauer[7,8,9], L. Pentericci[10], G. K. Miley[11], J. R. Allison[4], E. M. Sadler[12], P. Guillard[2], C. L. Carilli[6,13], M. Y. Mao[14,15], H. J. A. Röttgering[11], C. De Breuck[16], N. Seymour[17], B. Gullberg[18], D. Ceverino[1,19], P. Jagannathan[6], J. Vernet[16], B. T. Indermuehle[4]

[1] Centro de Astrobiología (Consejo Superior de Investigaciones Científicas – Instituto Nacional de Técnica Aeroespacial), Ctra de Torrejón a Ajalvir, km 4, 28850 Torrejón de Ardoz, Madrid, Spain
[2] Sorbonne Universités, Université Pierre et Marie Curie, Université Paris 6 et Centre National de la Recherche Scientifique, Unité Mixte de Recherche 7095, Institut d'Astrophysique de Paris, 98 bis Bd Arago, 75014 Paris, France
[3] Astro-Universidad Autónoma de Madrid, Unidad Asociada Consejo Superior de Investigaciones Científicas, Facultad de Ciencias, Campus de Cantoblanco, 28049, Madrid, Spain
[4] Commonwealth Scientific and Industrial Research Organisation Astronomy and Space Science, Australia Telescope National Facility, Post Office Box 76, Epping NSW, 1710, Australia
[5] Western Sydney University, Locked Bag 1797, Penrith South, NSW 1797, Australia
[6] National Radio Astronomy Observatory, Post Office Box 0, Socorro, NM 87801-0387, USA
[7] Instituto de Astrofísica de Canarias, E-38205 La Laguna, Tenerife, Spain
[8] Universidad de La Laguna, Departamento Astrofísica, E-38206 La Laguna, Tenerife, Spain
[9] Universität Wien, Institut für Astrophysik, Türkenschanzstraße 17, 1180 Wien, Austria
[10] Istituto Nazionale di Astrofisica, Osservatorio Astronomico di Roma, Via Frascati 33,00040 Monteporzio (RM), Italy
[11] Leiden Observatory, University of Leiden, P.O. Box 9513, 2300 RA Leiden, Netherlands
[12] Sydney Institute for Astronomy, School of Physics, University of Sydney, NSW 2006, Australia
[13] Cavendish Laboratory, 19 J. J. Thomson Ave., Cambridge, CB3 0HE, UK
[14] Joint Institute for Very Long Baseline Interferometry (VLBI) European Research Infrastructure Consortium, Postbus 2, 7990 AA, Dwingeloo, The Netherlands
[15] Jodrell Bank Observatory, University of Manchester, Macclesfield, Cheshire SK11 9DL
[16] European Southern Observatory, Karl Schwarzschild Strasse 2, 85748 Garching, Germany
[17] International Centre for Radio Astronomy Research, Curtin University, Perth, Australia
[18] Centre for Extragalactic Astronomy, Department of Physics, Durham University, South Road, Durham DH1 3LE, UK
[19] Zentrum für Astronomie der Universität Heidelberg, Institut für Theoretische Astrophysik, Albert-Ueberle-Str 2, 69120 Heidelberg

*Correspondence to: bjornemonts@gmail.com

**The largest galaxies in the Universe reside in galaxy clusters. Using sensitive observations of carbon-monoxide, we show that the Spiderweb Galaxy – a massive galaxy in a distant proto-cluster – is forming from a large reservoir of molecular gas. Most of this molecular gas lies between the proto-cluster galaxies and has low velocity dispersion, indicating that it is part of an**

**enriched inter-galactic medium. This may constitute the reservoir of gas that fuels the widespread star formation seen in earlier ultraviolet observations of the Spiderweb Galaxy. Our results support the notion that giant galaxies in clusters formed from extended regions of recycled gas at high redshift.**

The formation of the largest galaxies in the Universe is thought to be a two-stage process. For the last 10 Gyr, these giant galaxies have grown mostly by cannibalizing smaller galaxies *(1,2)*. However, computer simulations predict that in an earlier phase, lasting a few Gyr, their stars condensed directly out of large reservoirs of accreted gas *(3,4)*.

We present observational evidence for an extended gas reservoir fueling star formation in the massive Spiderweb Galaxy, MRC 1138-262, which is located in a proto-cluster at a redshift of $z = 2.161$ *(5-9)*. The Spiderweb Galaxy is not a single galaxy, but an aggregation of proto-cluster galaxies. They are embedded in a giant halo of atomic (neutral and ionized) hydrogen gas, which radiates Ly$\alpha$ emission across a region of ~200 kpc *(6)*. The central proto-cluster galaxy has a super-massive black hole at its core, which emits jets of relativistic particles visible in radio observations *(5)*. Observations suggest that the proto-cluster galaxies will eventually merge and evolve into a single, giant elliptical galaxy in the center of the cluster *(10)*. We therefore refer to the Spiderweb Galaxy as the entire region encompassed by the Ly$\alpha$ halo, and to the gas between the proto-cluster galaxies as the inter-galactic medium (IGM).

Earlier observations of line emission by carbon-monoxide revealed the presence of large amounts of molecular gas in the Spiderweb Galaxy *(11)*. Molecular gas is the raw fuel for the formation of stars, so observations of molecular gas give us insight into the processes driving the evolution of the distant Spiderweb Galaxy. We have obtained new, sensitive observations of $^{12}$CO (J=1→0) with the Australia Telescope Compact Array (ATCA, 90 hour exposure time) and the Karl G. Jansky Very Large Array (VLA, 8 hour exposure time) *(12)*. The ATCA observations were optimized for detecting low-surface-brightness emission from broadly distributed CO, with a 4.8″×3.5″ resolution. The VLA observations complement the ATCA data with a higher 0.7″×0.6″ resolution, sensitive to small-scale features but not to large-scale ones. Sampling these different spatial scales allows us to obtain a complete picture of the CO distribution, from the gas in the individual proto-cluster galaxies to that across the IGM.

Figure 1 shows that the CO emission in the ATCA data covers a region of ~70 kpc around the central radio galaxy. The CO (J=1→0) luminosity is $L'_{CO, ATCA} \sim (5.6\pm1.7) \times 10^{10}$ K km s$^{-1}$ pc$^2$. The high-resolution VLA data detect CO only on arcsecond scale within the central radio galaxy itself. Figure 2 shows that this central emission accounts for only one-third of the CO luminosity observed with the ATCA, or $L'_{CO, VLA} \sim (1.9\pm0.6) \times 10^{10}$ K km s$^{-1}$ pc$^2$. Thus, two-thirds of the CO detected with the ATCA, $(3.7\pm1.1) \times 10^{10}$ K km s$^{-1}$ pc$^2$, originates from gas outside the central radio galaxy. Figure S1 shows that the CO-emitting gas outside the central radio galaxy is spread on scales larger than the individual proto-cluster galaxies. This large-scale molecular gas cannot be imaged with our VLA data, which are far less sensitive to low-surface-brightness emission than the ATCA data (Fig. S1). Figure 3 shows that the bulk of this large-scale CO is not peaking co-spatially with the central radio galaxy or any of its brightest surrounding proto-cluster galaxies; the CO-emitting gas instead appears to be concentrated between the galaxies (supplementary online text). The velocity dispersion of the CO-emitting gas, $\sigma_{CO}$~220 km s$^{-1}$, is also much lower than that of the proto-cluster galaxies, $\sigma_{gal}$>1000 km s$^{-1}$

(Fig. 2) *(7)*. This agrees with the fact that our VLA data do not detect CO down to a 3σ limit of $L'_{CO} < 9 \times 10^9$ K km s$^{-1}$ pc$^2$ in any of the surrounding proto-cluster galaxies (supplementary online text). Based on Figs. 1, 2, 3 and S1, we therefore argue that our data show a molecular phase of the IGM embedded in the giant Lyα-emitting halo of the Spiderweb Galaxy.

There is no evidence that this large-scale molecular gas is rotating. Instead, the most blueshifted CO lies towards the southwest, while gas at increasingly higher velocities is distributed increasingly counter-clockwise around the radio galaxy (Fig. 3). The brightest CO peak lies in the region between two bright spots in the radio jet (Fig. 3E-F). This is consistent with recent detections of water (H$_2$O) emission along the jets in the Spiderweb galaxy, which imply rapid dissipation of the kinetic energy supplied by the relativistic jets *(13)*. Following recent models *(14,15)*, the peak in molecular CO emission along the jet also likely indicates that jet-induced gas cooling occurs within the IGM (supplementary online text).

*Hubble Space Telescope (HST)* imaging of extended rest-frame ultraviolet (UV) light in the IGM of the Spiderweb Galaxy has previously revealed ongoing star-formation across a region of ~70 kpc, with a star-formation rate (SFR) of SFR$_{IGM}$ ~ 142 M$_\odot$ yr$^{-1}$ *(16)*. The morphology of this UV-emitting region is similar to that of the CO emission (Fig. 3). The extended UV-light is most likely produced by young O- and B-type stars that formed in-situ within the IGM, because its restframe-UV color (as measured through the *HST* Advanced Camera for Surveys F475W and F814W filters) is bluer than that of the proto-cluster galaxies, and the extended UV-light is also not easily explained by nebular continuum, scattered light or synchrotron emission *(16)*. The estimated mass of molecular H$_2$ gas that is available to sustain this in-situ star-formation is M$_{H2,\ IGM}$ ~ (1.5±0.4) × 10$^{11}$ (α$_{CO}$/4) M$_\odot$, with α$_{CO}$ ≡ M$_{H2}$/$L'_{CO}$ the conversion factor between the CO luminosity and the molecular gas mass in the IGM *(17)*. We assume a value of α$_{CO}$ = 4 M$_\odot$(K km s$^{-1}$ pc$^2$)$^{-1}$, because the IGM of the Spiderweb Galaxy is likely to have a metallicity well below the solar value *(17)*, and because an independent estimate of the H$_2$ mass based on the dust emission *(18)* gives the same value (supplementary online text). Figure 4 shows that, with this available molecular gas mass, the in-situ star-formation in the IGM of the Spiderweb Galaxy falls on the relation between the surface-density of the SFR and the gas surface-density, also known as the Kennicutt-Schmidt relation *(19)* (supplementary online text). We therefore conclude that there is sufficient molecular gas available to fuel the in-situ star-formation within the IGM.

The molecular gas in the IGM of the Spiderweb Galaxy could sustain the current rate of in-situ star formation for a timescale $t_{depletion}$ ≡ M$_{H2}$/SFR$_{IGM}$ ~ 1.1(α$_{CO}$/4) Gyr. With our assumed value of α$_{CO}$, this could fuel the star formation until z~1.6. Therefore, even if the gas is replenished for another Gyr, our results are consistent with earlier predictions that the halo must stop forming stars by redshift 1, so that the stellar population has at least 5 Gyr time to age and reach the colors seen across the stellar halos of local central-cluster ellipticals *(10)*.

Our study of the Spiderweb Galaxy demonstrates that giant cluster-galaxies can grow their stellar mass in-situ out of very extended reservoirs of molecular gas, early in their formation process. The carbon and oxygen elements required to form the observed CO were made in the cores of stars. Therefore, the gaseous halo must have been polluted with recycled material that has been processed by previous episodes of star-formation and was subsequently expelled back into the IGM. Our CO results complement absorption-line studies of distant quasars, which infer the presence of large (≥100 kpc) halos of warm (T~10$^4$ K), metal-enriched gas *(20)*. Our results also support recent computer models on gas-infall into massive dark-matter halos of submillimeter-bright galaxies, which predict that stellar feedback from both galaxies and in-situ star-formation can enrich gas across ~200 kpc with dust and metals *(21)*. In addition to enrichment by star-formation, powerful radio jets may drag metals out of the host galaxy and far into the halo environment *(22)*. We therefore conclude that our observations have now identified the predicted cold baryon cycle that governs the early growth of massive cluster galaxies *(14,21)*.

## Acknowledgments

The research leading to these results has received funding from the People Programme (Marie Curie Actions) of the European Union's Seventh Framework Programme FP7/2007-2013/ under REA grant agreement nº 624351. We thank L. Colina and S. Arribas for useful discussions. BE and MVM received funding through grant AYA2012-32295 from the Spanish Ministerio de Economía y Competitividad (MINECO). HD's work was funded by MINECO under Ramón y Cajal program RYC-2014-15686. MYM acknowledges support from EC H2020-MSCA-IF-2014 660432. BG and DC acknowledge support from the ERC Advanced Grants DUSTYGAL 321334 and STARLIGHT 33177. The Australia Telescope is funded by the Commonwealth of Australia for operation as a National Facility managed by CSIRO. The National Radio Astronomy Observatory is a facility of the National Science Foundation operated under cooperative agreement by Associated Universities, Inc. The data reported in this paper are tabulated in the Supplementary Materials and archived at http://atoa.atnf.csiro.au/ under project codes C2052 and C2717 (ATCA), and at https://archive.nrao.edu/archive/advquery.jsp under project code 14B-160 (VLA).

## Supplementary Materials

www.sciencemag.org
Materials and Methods
Supplementary Text
Fig. S1
Tables S1, S2, S3
References (25-61)

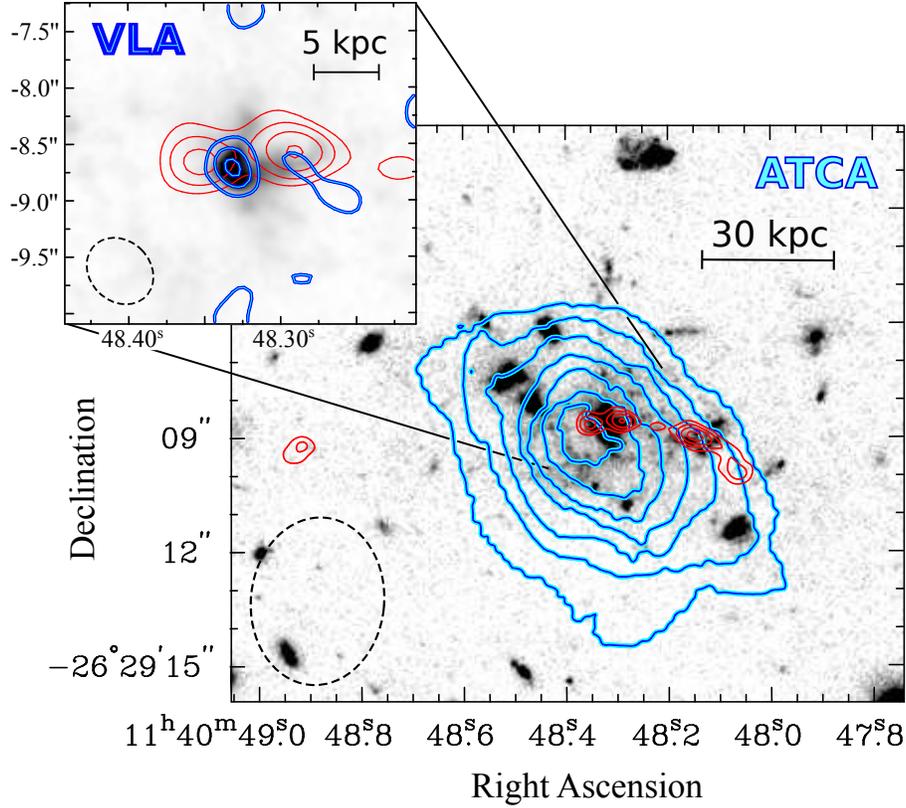

**Fig. 1. Molecular gas on multiple scales in the Spiderweb Galaxy.** $^{12}$CO (J=1→0) total-intensity contours from the Australia Telescope Compact Array (light blue) overlaid onto a negative grey-scale *Hubble Space Telescope* image taken with the Advanced Camera for Surveys through the combined F475W and F814W filters (*HST* image ©AAS, reproduced with permission *(6)*). Contour levels: 0.020, 0.038, 0.056, 0.074, 0.092, 0.110, 0.128 Jy beam$^{-1}$ × km s$^{-1}$. Red contours show the 36 GHz radio-continuum from our Very Large Array (VLA) data at 0.20, 0.43, 0.80, 1.40 mJy beam$^{-1}$. The top-left inset shows the CO (J=1→0) total-intensity contours from the VLA (dark blue) at 2.8σ, 3.5σ, 4.2σ, with σ=0.019 Jy beam$^{-1}$ × km s$^{-1}$. No negative contours are visible at this level around the central radio galaxy in the VLA data, likely because underlying large-scale flux skews the noise to slightly more positive values (Fig. S1). The uncertainty in the astrometry of the *HST* image is ~0.3″. The two inner radio-continuum components (red contours in the top-left inset) likely trace the two-sided base of the radio jets. The dashed ellipses in the bottom left-hand corners visualize the beam-size at full width of half the maximum intensity (FWHM). Coordinates are given in epoch J2000.

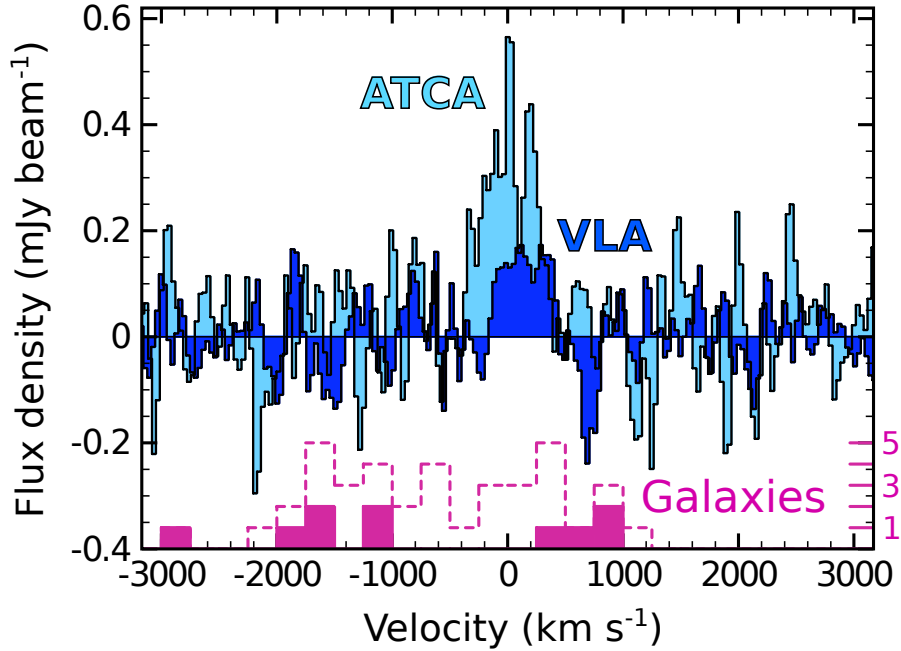

**Fig. 2. Spectra of the CO-emitting gas.** $^{12}$CO (J=1→0) spectra at the location of the peak emission in Fig. 1. The light and dark blue spectra show the data taken with the Australia Telescope Compact Array (ATCA) and the Karl G. Jansky Very Large Array (VLA), respectively. The ATCA spectrum was obtained by tapering the data to a projected baseline-length of ~200m to ensure that we recover all the CO emission (Fig. S1 and *(12)*). The VLA spectrum shows the CO emission at the highest spatial resolution. The histogram outlined with a dashed magenta line shows the velocity distribution of the proto-cluster galaxies *(7)*. The solid magenta histogram only includes galaxies that lie within the CO-emitting region, with velocities derived from the rest-frame optical emission lines [O II], [O III] and Hα *(7)*. The corresponding number of galaxies per bin is indicated in magenta on the right axis. Velocities are with respect to z = 2.161 *(12)*.

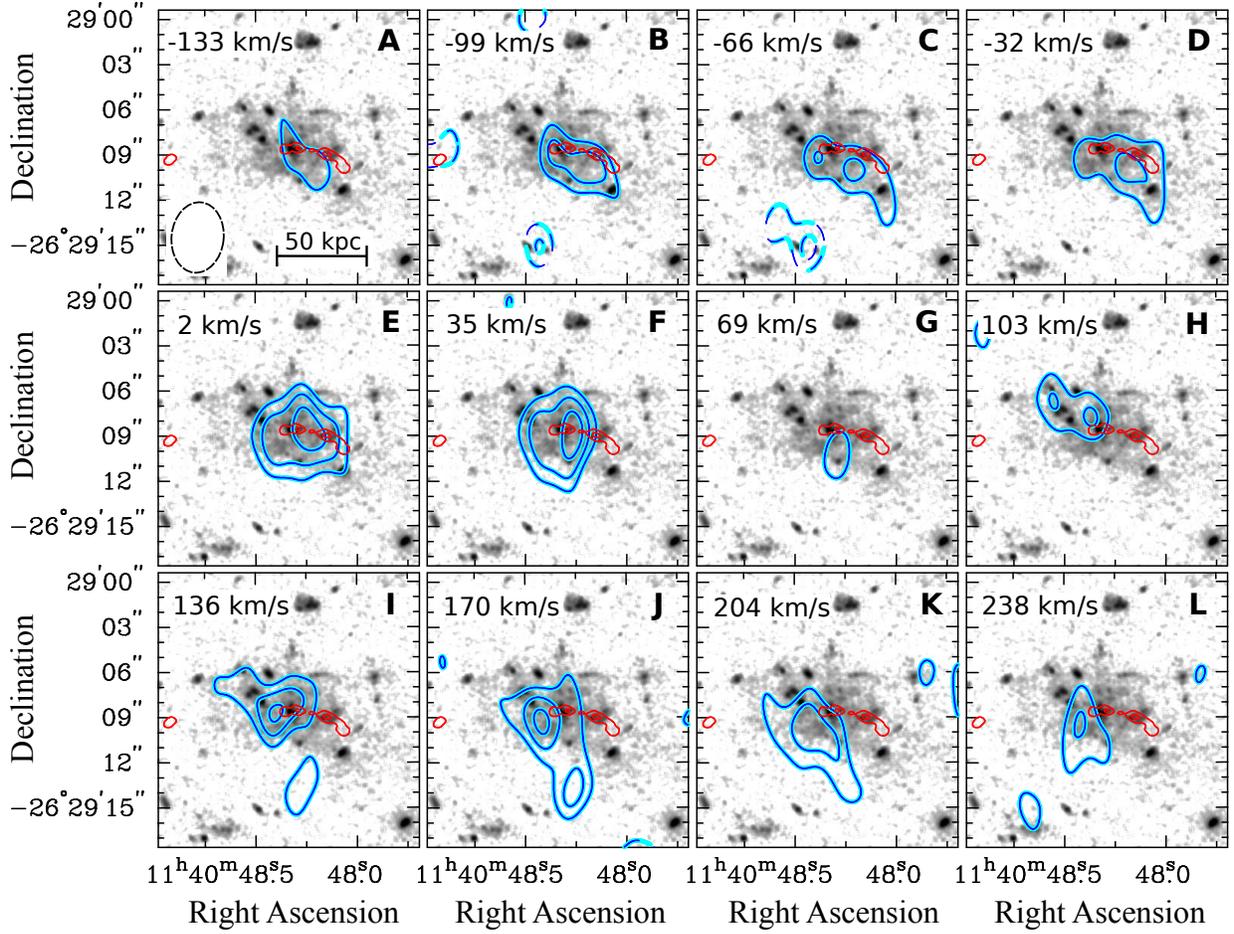

**Fig. 3. Distribution and kinematics of the CO-emitting gas.** Maps of $^{12}$CO (J=1→0) detected with the Australia Telescope Compact Array at different velocities, increasing in each panel A-L. Blue CO contour levels are at -3.5σ, -2.5σ (both dashed), 2.5σ, 3.5σ, 4.5σ (all solid), where σ=0.085 mJy beam$^{-1}$. The background grey-scale image was taken with the *Hubble Space Telescope* Advanced Camera for Surveys through the F475W-filter. This *HST* image is placed on a logarithmic scale to show the extended rest-frame ultraviolet light from in-situ star formation discovered in earlier work *(16)*. We downloaded these data from the *HST* Legacy Archive and re-processed them by applying a 5×5-pixel boxcar-smooth to highlight the low surface-brightness ultraviolet emission. Red contours show the radio-continuum source from Fig. 1. The beam-size of the CO data is visualized with the dashed ellipse in panel A. The positional accuracy of the CO peaks is ~0.5″ for a 4.5σ signal (supplementary online text). The CO signal is not entirely independent across adjacent panels because of the applied Hanning smooth *(12)*. Velocities are with respect to *z* = 2.161 *(12)*. Coordinates are given in epoch J2000.

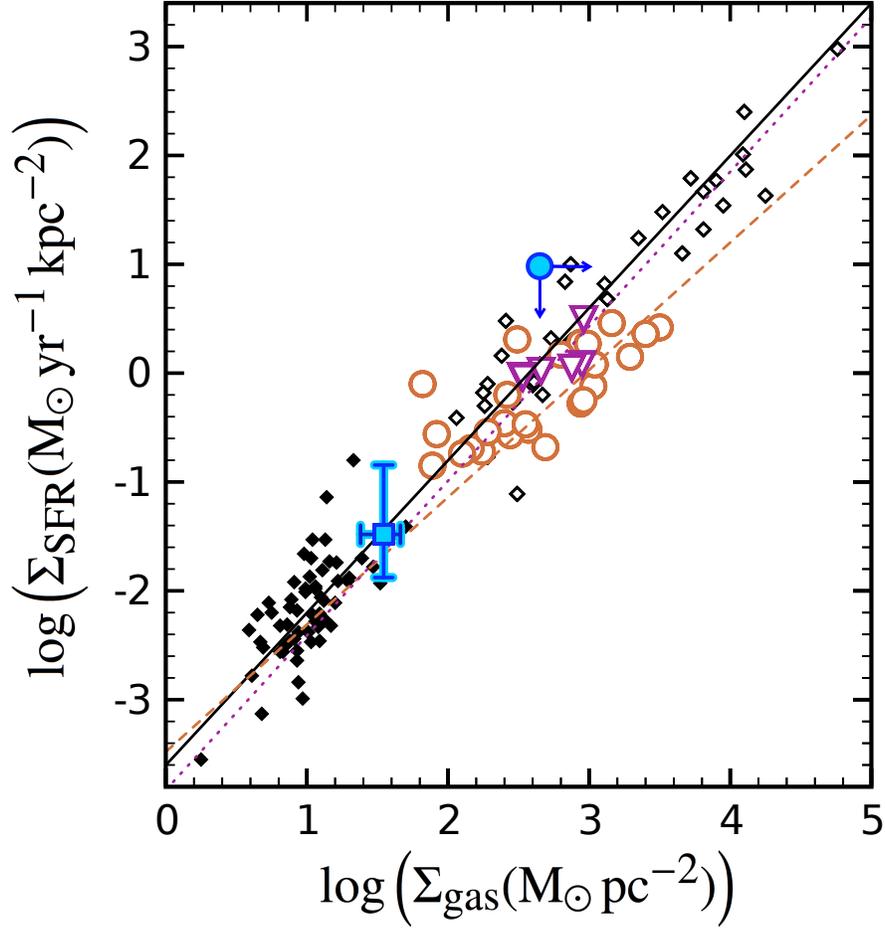

**Fig. 4. The molecular star-forming gas in the IGM on the Kennicutt-Schmidt relation** *(19)*. The surface-density of the star-formation rate ($\Sigma_{SFR}$) is plotted as a function of the molecular gas mass surface-density ($\Sigma_{gas}$). The solid blue square represents the in-situ star-formation across the IGM of the Spiderweb Galaxy. The solid blue circle shows the limits for the central radio galaxy from our high-resolution VLA data for $\alpha_{CO} \geq 0.8$ M$_\odot$(K km s$^{-1}$ pc$^2$)$^{-1}$. For limits and error-calculations, see supplementary online text. Black diamonds represent local spiral galaxies (solid) and circum-nuclear starbursts (open) *(19)*. Brown circles and purple triangles represent star-forming galaxies at $1.0 \leq z \leq 3.5$ *(23)* and $z \sim 1.5$ *(24)*, respectively. The lines represent the best fit to the different sample (black solid *(19)*, brown dashed *(23)* and purple dotted *(24)*).

# *Science* (AAAS)

# Supplementary Materials for

Molecular Gas in the Halo Fuels the Growth of a Massive Cluster Galaxy at High Redshift

B. H. C. Emonts, M. D. Lehnert, M. Villar-Martín, R. P. Norris, R. D. Ekers, G. A. van Moorsel, H. Dannerbauer, L. Pentericci, G. K. Miley, J. R. Allison, E. M. Sadler, P. Guillard, C. L. Carilli, M. Y. Mao, H. J. A. Röttgering, C. De Breuck, N. Seymour, B. Gullberg, D. Ceverino, P. Jagannathan, J. Vernet, B. T. Indermuehle

**This PDF file includes:**

Materials and Methods
Supplementary Text
Fig. S1
Tables S1 – S3
References 25-61

## Materials and Methods

We performed the CO observations during April 2011 – Feb 2015 with the Australia Telescope Compact Array (ATCA) *(25)* and the Karl G. Jansky Very Large Array (VLA) *(26)*, see Table S1. The observing band was centered on the redshifted (z=2.161) $^{12}$CO (J=1→0) line at 36.46 GHz (8.2 mm) and covered an effective width of 2 GHz at the ATCA and 1 GHz at the VLA. We used raw channels with a width of 1 MHz, or ~8.5 km s$^{-1}$. For phase calibration we obtained short (1-2min) observations of PKS 1124-186, PKS 1143-287 or PMN J1209-2406 every 3.5 to 12 min, the cadence depending on array configuration and weather conditions. Bandpass calibration was performed with a 5 min observation of PKS 1124-186 at least once per hour. Pointing corrections were obtained by self-calibrating the antennas through a standard pointing pattern on the calibrator source each time the telescope moved >15° away from the sky-position of the last pointing. Flux densities were calibrated using Mars, PKS 1934-638 and the ultra-compact H II region G309 (with flux density bootstrapped from Uranus *(27)*) at the ATCA, and 3C 286 at the VLA. See Table S1 for details.

We reduced the ATCA data using Miriad *(28)* and the VLA data using the Common Astronomy Software Applications (CASA) *(29)*. After applying the bandpass, phase and flux calibration to the individual data sets, we used the non-variable continuum-flux of the radio source in the Spiderweb Galaxy to determine the accuracy of the flux calibration. We verified that the flux scale between all calibrated data sets is the same to within a typical 30% accuracy for absolute flux calibration at both the ATCA and the VLA. We imaged the radio continuum in the VLA data (red contours in Fig. 1) using a robust +0.3 weighting with a resolution of 0.5″ × 0.4″ at position angle (PA) 78°. We then separated the continuum from the line-emission in the complex visibility (u,v) domain by fitting a straight line to the line-free channels. The line-data were then Fourier transformed using natural weighting. We translated all velocities into the barycentric reference-frame (using the optical convention) with respect to $z$ = 2.161. We then binned the ATCA and VLA data to 34 km s$^{-1}$ channels and subsequently Hanning smoothed these data to an effective velocity resolution of 68 km s$^{-1}$. The resulting noise levels are 0.085 mJy beam$^{-1}$ per channel for the ATCA and 0.080 mJy beam$^{-1}$ per channel for the VLA data. More details of the line-data properties are shown in Table S2. To create a total intensity image from the ATCA data, we used a mask to ensure that we sum all signal in the regions where CO is present *(30)*. We made this mask by tapering the original line-data to include only baselines up to 31 kλ and subsequently smoothing these data to a beam-size of 6.8″ × 5.8″. We then blanked out all signal in the original data set in those pixels that have a corresponding value below 2.5σ in the mask. We summed the remaining signal in this data set to create a total intensity image of the CO emission. We stress that the mask was only used to create the total intensity image shown in Fig. 1, and not for the data products shown in Figs. 2, 3 and S1. For the weaker VLA signal, we made a total intensity image by summing the channels with CO.

Figure S1 shows that the ATCA and VLA data are sensitive to tracing CO emission on very different spatial scales, and that their flux-density scales are accurately calibrated. We note that the VLA data do recover the ATCA flux on the shortest baselines, but the corresponding noise level on these short baselines is too high to image the most extended CO emission with these VLA data. We could not reliably combine the ATCA and VLA data sets, because there are intrinsic uncertainties in combining the data obtained at two telescopes and reduced with different software. In addition, the (u,v) coverage of the combined ATCA and VLA data was not optimal, because we experienced the poorest weather conditions during the observations in the most extended ATCA (1.5A) configuration and the most compact VLA (DnC) configuration (for details on the array configurations, see electronic link in Table S1). This resulted in a limited sensitivity on intermediate (~1 km) baselines.

Table S3 summarizes the CO properties that we derived from the ATCA and VLA data. The coordinates of the CO emission in the VLA data were obtained from the total intensity image of Fig. 1, while the CO (J=1→0) intensity ($I_{CO}$) and luminosity ($L'_{CO}$) were obtained by integrating the signal across the line profiles of Fig. 2. The uncertainty in $I_{CO}$ and $L'_{CO}$ is dominated by the uncertainty in absolute flux calibration and therefore assumed to be 30%. The full width at half the maximum intensity (FWHM$_{CO}$), the CO velocity dispersion ($\sigma_{CO}$ = FWHM$_{CO}$/2.35), and CO redshift ($z_{CO}$) were estimated by fitting a Gaussian function to the line profiles of Fig. 2. The ATCA CO luminosity that we report in this paper is in agreement with previous ATCA work *(11)*. However, our new data do not detect the CO-features within the Spiderweb Galaxy that were previously reported as tentative, namely the 3σ tentative blue peak towards the southwest and the tentative red tail northeast of the radio galaxy *(11)*. Our new data do confirm the previously reported tentative detection of the proto-cluster galaxy HAE229 about 250 kpc west of the Spiderweb Galaxy *(11)*. We determined the size of the CO-emitting region from the channel maps of Fig. 3. We convolved the delta-functions that correspond to the peak emission per channel in Fig. 3 with the synthesized beam and subtracted this from the signal in each channel. The residual CO signal extends across a diameter of ~70 kpc at a significance level of 3σ. From the high-resolution VLA data, we derived a 3σ upper limit to the CO luminosity in the individual proto-cluster galaxies of $L'_{CO} < 9 \times 10^9$ K km s$^{-1}$ pc$^2$, assuming that $I_{CO} < 3\sigma \Delta v \sqrt{(FWHM_{CO}/\Delta v)}$, where FWHM$_{CO}$ = 400 km s$^{-1}$ is the assumed CO velocity width and $\Delta v$ = 68 km s$^{-1}$ the effective velocity resolution of our data.

Throughout the paper we translated the observed CO results into physical properties using the following cosmological parameters: $H_0$ = 71 km s$^{-1}$ Mpc$^{-1}$, $\Omega_M$ = 0.27 and $\Omega_\Lambda$ = 0.73. These parameters correspond to an angular distance scale of 8.4 kpc arcsec$^{-1}$ and a luminosity distance of $D_L$ = 17309 Mpc at the redshift of the CO emission line *(11,31)*.

## Supplementary text

1. Positional accuracy CO features

Because the CO signal is kinematically resolved and the CO-peaks in each channel have a signal of 4-5σ (Fig. 3), the distribution of the gas can be studied by the spatial shift of the signal-peak on scales significantly smaller than the beam-size. The positional accuracy of the CO emission-line peaks in the channel maps of Fig. 3 is then *(32)*:

$\delta\theta_{rms} \sim 0.5 \langle\Theta_{beam}\rangle (S/N)^{-1} \sim 0.5''$,  (S1)

for $\Theta_{beam}$ = 4.5″, the FWHM of the synthesized beam, and S/N = 4.5, the signal-to-noise ratio. We cannot rule out that multiple CO peaks that have the same velocity and are located within one beam blend into the CO features that we observe in the velocity channels of Fig. 3. Still, we are confident that such signal would not be dominated by CO that is located within individual proto-cluster galaxies, because the velocities of the galaxies are in most cases very different from those of the molecular gas *(7)*.

The uncertainty in the absolute astrometry of our CO data is small enough not to influence these results. Given that the most distant phase calibrator that we used (PKS 1124-186) is 8.17° away from our target, the uncertainty in absolute astrometry due to the phase error $\delta\phi_{bas}$

~ $(2\pi/\lambda)(\delta \boldsymbol{B} \cdot \Delta \mathbf{k})$ [where $|\delta \boldsymbol{B}| \sim 1$mm the typical error in the baseline-length and $|\Delta \mathbf{k}| = 8.17°$ the distance to the phase calibrator] is *(32)*:

$$\delta\theta_{bas} = (\delta \boldsymbol{B} \cdot \Delta \mathbf{k})/\mathrm{B} \approx (\delta\phi_{bas}/2\pi) \langle \Theta_{beam} \rangle \sim 0.1'', \quad (S2)$$

where B is the typical baseline-length. This is consistent with the fact that the 36 GHz radio-continuum source in our VLA data perfectly aligns with the 8.2 GHz continuum source from *(5)*.

2. Jet-induced gas cooling in a multi-phase medium

Recent models of gas in local galaxy clusters suggest that even if heating by the radio jet is sufficient to balance the overall cooling of the hot thermal cluster gas, the energy injection causes entropy fluctuations that lead to local thermal instability and gas cooling *(14,15,33,34)*. Previous observations of the Spiderweb Galaxy showed that it is surrounded by extended emission of Lyα and X-rays *(5,6,35)*. The Lyα emission indicates that there is warm, atomic (neutral and ionized) gas with a temperature of ~$10^4$ K over a scale of ~200 kpc around the radio galaxy, with enhancements in emission along the radio jet *(5,6)*. The thermal X-ray emission indicates the IGM contains hot (T~$10^{7-8}$ K), low density (n~0.05 cm$^{-3}$) gas with a relatively long cooling time (~$10^9$ yr) *(35)*. Our detection of extended molecular gas in the IGM of the Spiderweb Galaxy indicates it has a multiphase medium likely spanning more than five orders of magnitude in both density and temperature. While the masses and cooling rates of the hot thermal gas and warm atomic medium are sufficient to maintain the relatively cooler molecular gas *(35)*, the enhancements of the X-ray, Lyα, water *(13)*, and CO emission along the radio jet are likely indications of both the heating and the cooling the radio jets induce as they propagate outwards from the super-massive black hole. The multiphase nature of the IGM of the Spiderweb Galaxy is similar to what has been seen around the massive central galaxies in nearby clusters, like the Perseus Cluster *(36-40)*. The centers of nearby clusters have filaments of molecular gas, despite the undeniable evidence for jet-induced heating of gas in these systems *(39,40)*. The fraction of gas that is cooled rather than heated by the propagating radio jets and surrounding X-ray emitting gas depends crucially on the structure of the warm and cold gas clouds and the characteristics of the magnetic fields *(41)*. For the Spiderweb Galaxy, deriving the detailed relationship between the gas phases is beyond the capabilities of existing telescopes. We note that some of the dense, cool clouds that form in the IGM in this way may become self-gravitating and collapse to form stars *(42)*, but we caution that our theoretical understanding of the multiphase nature of gas in proto-clusters and how stars form is rudimentary.

3. Molecular gas mass

The intensity of the radiation emitted by CO molecules through the lowest rotational transition, $^{12}$CO (J=1→0), is much less affected by excitation conditions of the gas than the emission of the higher-J transitions. Therefore, CO (J=1→0) provides a reliable estimate of the overall molecular gas mass, including any widespread and sub-thermally excited molecular component *(43)*. To translate CO (J=1→0) luminosities into molecular gas masses, a conversion factor, $\alpha_{CO} \equiv M_{H2}/L'_{CO}$ is needed *(17)*. To estimate the molecular gas mass within the IGM, we assume a conversion factor of $\alpha_{CO} \sim 4$ M$_\odot$ (K km s$^{-1}$ pc$^2$)$^{-1}$. This value is close to the value for molecular clouds in the Milky Way *(44)*, as well as normal star-forming galaxies at high redshifts *(23,24)*. Choosing this value may yield a conservative estimate for the total molecular gas mass, because the IGM of the Spiderweb Galaxy may have a relatively low metallicity compared to gas within galaxies. If we consider that the IGM in low-z clusters has a metallicity of ~0.3 times the solar value $Z_\odot$ *(45,46)*, and simulations show that 35-60% of these metals were injected by z~2 *(47)*, then metallicities of ~0.1-0.2 $Z_\odot$ in the IGM of the Spiderweb Galaxy are plausible. Such low metallicities may suggest that $\alpha_{CO} > 4$ *(17)*. On the other hand, metallicities close to the solar value have also been reported for the halo of a quasar at redshift 2.4 *(48)*. This perhaps could imply that the IGM has a conversion factor closer to the $\alpha_{CO} \sim 0.8 - 1.6$ $M_\odot$ (K km s$^{-1}$ pc$^2$)$^{-1}$ typically assumed for high-*z* submillimetre galaxies and some star-forming galaxies *(49,50)*. The conversion factor also depends on other properties, such as the intensity of the stellar radiation field and its extinction by dust *(17,51-53)*. These properties are not yet well understood even within galaxies, let alone in the gaseous medium outside of galaxies. Keeping these possible systematic uncertainties in mind, we estimate from the CO (J=1→0) emission that the IGM of the Spiderweb Galaxy contains a molecular gas mass of $M_{H2, IGM} \sim 1.5 \pm 0.4 \times 10^{11}$ ($\alpha_{CO}/4$) $M_\odot$.

We can derive an independent gas-mass estimate from the thermal dust emission seen in earlier sub-millimeter observations, whose morphology appears to roughly follow that of the CO emission *(18,54)*. If we assume that, just like for the CO line emission, also two-thirds of the dust *(18)* is spread throughout the IGM, this corresponds to $M_{dust, IGM} \sim 3.1 \times 10^8$ $M_\odot$ if the excitation conditions of the dust are uniform. When scaling the Milky Way's oxygen abundance and dust content to a dusty medium with 0.2 $Z_\odot$, we expect a dust-to-gas mass ratio of ~0.2% *(55)*. This corresponds to $M_{gas, IGM} \sim 1.5 \times 10^{11} M_\odot$, or similar to what we estimated using the CO line luminosity and $\alpha_{CO} = 4$ $M_\odot$ (K km s$^{-1}$ pc$^2$)$^{-1}$.

We caution that the intrinsic luminosity of the CO line-emission, and thus also the molecular gas mass, may be underestimated in molecular gas at high redshifts because of the increased temperature of the Cosmic Microwave Background (CMB). The CMB at z=2.16 has a temperature of $T_{CMB} = 8.6$ K. The CMB provides an additional source of heating, as well as a strong background continuum against which the CO emission needs to be measured. Both effects lead to a reduction in the measured CO luminosity compared to the emitted luminosity for constant molecular gas mass *(56)*. Theoretical work shows that for low gas temperatures (T ~ 20 K), due to these effects, the measured CO (J=1→0) luminosity can underestimate the intrinsic CO (J=1→0) luminosity at z~2 by as much as ~30% across a wide range of densities and excitation conditions *(56,57)*. If the gas has higher temperatures, the effect is less important. This effect of the CMB is therefore not expected to exceed the uncertainty of our flux calibration. Our data do not constrain the temperature of the molecular gas, therefore we do not correct $L'_{CO}$ for this effect. In any event, it shows that our $L'_{CO}$ estimates are conservative. The CO-emitting gas that we detect with the VLA within the central radio galaxy is in close proximity to the AGN and central starburst, and is thus likely to have a higher average gas temperature than the extended molecular gas across the IGM. This means that also the 2:1 ratio of $L'_{CO}$ between the molecular gas in the IGM and that in the central radio galaxy is likely a lower limit.

4. Kennicutt-Schmidt relation

The Kennicutt-Schmidt (KS) relation is an empirical relation between the gas surface-density and the surface-density of the star-formation rate, which holds over a wide range of galaxy types, from circum-nuclear starbursts in local galaxies to galaxy-wide star formation at high redshifts *(19)*:

$$\Sigma_{SFR} = (2.5 \pm 0.7) \times 10^{-4} \, (\Sigma_{gas}/[1 \, M_\odot \, pc^{-2}])^{1.4 \pm 0.15} \, M_\odot \, yr^{-1} \, kpc^{-2} \quad (S3)$$

The extended UV-light in the Spiderweb Galaxy occurs across an area of ~4200 kpc² *(16)*. With a dust-corrected $SFR_{IGM} \sim 142 \, M_\odot \, yr^{-1}$, this corresponds to an average $\Sigma_{SFR, \, IGM} \sim 0.033^{+0.114}_{-0.019} \, M_\odot \, yr^{-1} \, kpc^{-2}$. The lower-limit uncertainty in this estimate reflects the case of the observed $SFR_{IGM} \sim 57 \, M_\odot \, yr^{-1}$ before dust correction *(16)*. The upper-limit estimate reflects the integrated intrinsic $SFR_{IGM} \sim 1400 \, M_\odot \, yr^{-1}$ as measured across the Spiderweb Galaxy with low-resolution *Spitzer* and *Herschel* observations *(58,59)*, but following other work that indicates that at most ~44% of this star formation occurs in the IGM outside the proto-cluster galaxies *(16)*. Based on the KS-relation (equation S3), when taking $SFR_{IGM} \sim 142 \, M_\odot \, yr^{-1}$, we expect a gas mass surface-density of $\Sigma_{gas, \, IGM} \sim 33^{+32}_{-13} \, M_\odot \, pc^{-2}$. Because the spatial resolution of the UV-imaging from *HST* is higher than that of our CO data, we assume that the CO emission that is found outside the central radio galaxy is spread across the same region as the extended UV-light. This results in an observed average $H_2$ surface-density within the IGM of $\Sigma_{H2, \, IGM} \sim 35 \pm 11 \, M_\odot \, pc^{-2}$. This estimate is in agreement with the predicted value from the KS-relation for normal star forming galaxies at both low- and high-redshift, as shown in Fig. 4.

For the central radio galaxy (blue triangle in Fig. 4), we derive a lower limit for the gas surface-density from our high-resolution VLA data. We assume $\alpha_{CO} \geq 0.8 \, M_\odot (K \, km \, s^{-1} \, pc^2)^{-1}$, which is a lower-limit value of $\alpha_{CO}$ observed for ultra-luminous infrared galaxies and high-z star-forming galaxies *(49,50,60)*. This results in $M_{H2,RG} \geq 1.5 \times 10^{10} \, M_\odot$. We estimate the corresponding upper limit of $\Sigma_{SFR, \, RG}$ by using $SFR_{RG} < 440 \, M_\odot \, yr^{-1}$, as previously estimated from the 740 GHz rest-frame continuum in ALMA data that have comparable resolution to our VLA data *(13)*. These values result in $\Sigma_{gas,RG} \geq 450 \, M_\odot \, pc^{-2}$ and $\Sigma_{SFR, \, RG} < 13 \, M_\odot \, yr^{-1} \, kpc^{-2}$ over the area covered by the solid angle of the 0.7″×0.6″ VLA beam.

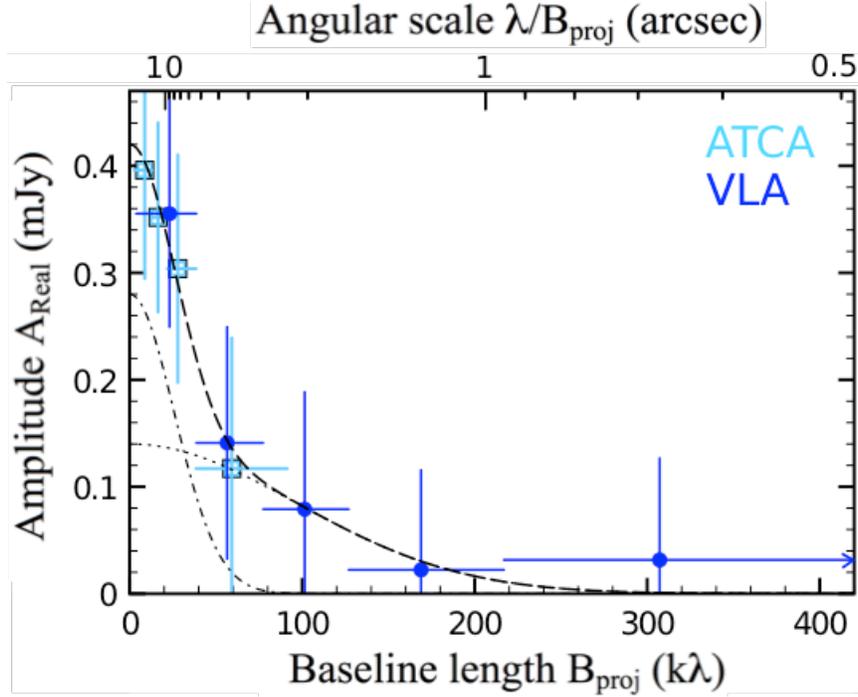

**Fig. S1. CO emission in the data from the Australia Telescope Compact Array and Very Large Array across different spatial scales.** Visibility-amplitudes of the CO (J=1→0) emission are plotted as function of the projected baseline length in the ATCA (light-blue squares) and VLA (dark-blue circles) emission-line data sets. The projected baseline length is defined as the uv-distance, $\sqrt{u^2+v^2}$, in units of 1000 times the wavelength. The amplitude plotted on the y-axis is the real part of the complex interferometer visibility for the CO-signal per raw channel, when averaged across the velocity range -100 to +200 km s$^{-1}$ and over all baselines within the uv-range covered by the horizontal error bars. The ±1σ uncertainty in the amplitude of the averaged signal in each bin is indicated with the vertical error bars. The dashed line shows a non-unique model in which we superimpose a Gaussian distribution with FWHM = 3″ (dash-dotted line) and a Gaussian distribution with FWHM = 0.8″ (dotted line) with an intensity-ratio of 2:1. We note that this plot of the real parts of the vector-averaged amplitudes may not fully reveal the true extent of the velocity-integrated CO emission, because only the CO signal centered on the position of the radio galaxy is included with full amplitude. Nevertheless, this plot shows the data at a higher signal-to-noise level than the power-spectrum, therefore we use it to illustrate that the ATCA and VLA data are sensitive to vastly different physical scales and that the flux calibration between the ATCA and VLA data is consistent.

**Table S1.** Overview of observations

| Configuration[*] | Observing dates | On-source exposure time (h) | Calibrators (P=phase, B=bandpass, F=flux) |
|---|---|---|---|
| ATCA H75[†] | 26,27 Sept 2011 | 5 | PKS 1124-186 (P+B), Mars (F) |
| ATCA H168[†] | 18,19 Aug 2011 13-17 Mar 2012 | 18 | PKS 1124-186 (P+B), Mars (F) |
| ATCA H214 | 21-31 Mar 2013 | 30 | PKS 1124-186 (P+B), G309 (F) |
| ATCA 750A/D | 17-22 Feb 2015 | 26 | PKS 1143-287 (P), PKS 1124-186 (B), PKS 1934-638 (F) |
| ATCA 1.5A[‡] | 14-18,29 Oct 2014 | 11 | PKS 1143-287 (P), PKS 1124-186 (B), PKS 1934-638 (F) |
| VLA DnC | 3 Oct 2014 | 1.7 | PMN J1209-2406 (P), PKS 1124-186 (B), 3C 286 (F) |
| VLA CnB | 11,12,15,17 Jan 2015 | 6.0 | PMN J1209-2406 (P), PKS 1124-186 (B), 3C 286 (F) |

Table Notes:
[*] See https://www.narrabri.atnf.csiro.au/operations/array_configurations/configurations.html and https://science.nrao.edu/facilities/vla/proposing/configpropdeadlines for details on the array configurations.
† The ATCA observations in the H75 and H168 configuration were part of previous work *(11,61)*.
‡ The ATCA observations in the 1.5A-configuration were done during mediocre weather conditions and the corresponding data-quality for baselines ≥1 km was poor. We therefore only included baselines up to ~800m for the 1.5A-configuration data. In addition, due to the relatively limited observing time in 1.5A-configuration, discarding these longest baselines also resulted in a much more uniformly covered uv-space.

**Table S2.** Overview of line-data properties

|  | ATCA | VLA |
|---|---|---|
| On-source time (hours) | 90 | 8 |
| Minimum baseline length (m) | 31 | 35 |
| Maximum baseline length (m) | 800 | 3,400[*] |
| Minimum baseline length (k$\lambda$) | ~2.7 | ~3.2 |
| Maximum baseline length (k$\lambda$) | ~92 | ~630 |
| Spatial resolution[†] (arcsec) | 4.8 × 3.5 | 0.7 × 0.6 |
| Position angle synthesized beam (°) | -6 | 45 |
| Velocity coverage (km s$^{-1}$) | 17,000 | 8,500 |
| Spectral resolution (km s$^{-1}$) | 68 | 68 |
| Channel width (km s$^{-1}$) | 34 | 34 |
| Root-mean-square noise (mJy beam$^{-1}$ chan$^{-1}$) | 0.085 | 0.080 |

Table Notes:
* Baselines on the extended north-south spur reach out to 11 km, but the corresponding projected baseline lengths are significantly shorter given the southern declination of -26° for the Spiderweb Galaxy.
† Measured at the half-power point of the synthesized beam, using natural weighting.

**Table S3.** Derived CO properties

|  | ATCA | VLA |
|---|---|---|
| Right Ascension CO (h:m:s)* | - | 11:40:48.33 ± 0.01s |
| Declination CO (°:′:″)* | - | -26:29:08.7 ± 0.1″ |
| $I_{CO}$ (Jy beam$^{-1}$ × km s$^{-1}$) | 0.24 ± 0.07 | 0.08 ± 0.03 |
| $L'_{CO}$ (×10$^{10}$ K km s$^{-1}$ pc$^2$) | 5.6 ± 1.7 | 1.9 ± 0.6 |
| FWHM$_{CO}$ (km s$^{-1}$) | 518 ± 52 | 418 ± 56 |
| $\sigma_{CO}$ = FWHM$_{CO}$/2.35 (km s$^{-1}$) | 220 ± 22 | 178 ± 24 |
| $z_{CO}$ | 2.1613 ± 0.0001 | 2.1617 ± 0.0003 |

Table Notes:
* Based on the location of the unresolved CO emission in the VLA data. The CO emission detected in the ATCA data is spatially extended.